\documentclass[aps,pra,twocolumn]{revtex4}
\usepackage{newlfont}
\usepackage{amssymb}
\usepackage{amsfonts}
\usepackage{amsmath}
\usepackage{graphicx}
\usepackage{bm}
\newcommand{\ket}[1]{|#1\rangle}
\newcommand{\bra}[1]{\langle#1|}

\begin{document}




\title{Usefulness of classical communication for local cloning of entangled states}

\author{Rafa{\l} Demkowicz-Dobrza\'nski\(^{1}\), Maciej Lewenstein\(^{2,3}\),
Aditi Sen(De)\(^{3,4}\), Ujjwal Sen\(^{3,4}\), Dagmar Bru\ss\(^{5}\)}

\affiliation{\(^{1}\)Center for Theoretical Physics, Polish Academy of Sciences
Aleja Lotnik{\'o}w 32/44, 02-668 Warszawa, Poland\\
\(^{2}\)ICREA and ICFO-Institut de Ci\`encies Fot\`oniques, E-08860  Castelldefels (Barcelona), Spain\\
\(^{3}\)Institut f\"ur Theoretische Physik, Universit\"at Hannover, D-30167 Hannover,
Germany\\
\(^{4}\)ICFO-Institut de Ci\`encies Fot\`oniques, E-08860  Castelldefels (Barcelona), Spain\\
\(^{5}\)Institut f\"ur Theoretische Physik III, Heinrich-Heine-Universit\"at
D\"usseldorf, 40225 D\"usseldorf, Germany}


\begin{abstract}
We solve the problem of the optimal cloning of pure entangled two-qubit states with a fixed degree of entanglement 
using local operations and classical communication.
We show, that amazingly, classical communication between the parties can improve the fidelity of local cloning
\emph{if and only if} the initial entanglement is higher
than a certain \emph{critical} value. It is completely useless for weakly entangled states. 
We also show that bound entangled states with positive partial transpose 
are not useful as a resource to improve the best local cloning fidelity. 

\end{abstract}

\maketitle

\section{Introduction}

Quantum  information processing (QIP) and engineering in quantum mechanical systems has always been regarded 
on two levels of complexity. First, one can consider quantum nature of single systems with regard to global operations. 
The second, and usually more complex, level is to consider composite systems with regard to local operations. 
The examples of such duality  are numerous: Quantumness \cite{pakhi} vs. entanglement \cite{lagaan}, 
global vs. local accessible information \cite{sundar, royal-bengal}, global vs. local dense coding 
\cite{chandrabindu, bisarga}, etc. In this paper, we discuss for the first time the duality between global and local cloning 
and find optimal local cloner for pure states of two qubits.


The impossibility of perfect cloning \cite{wootters} is a fundamental law
that Nature imposes on QIP.
Although perfect quantum
cloning is forbidden, 
approximate quantum cloning is possible \cite{buzek1996, cloning, werner1998, cloninf}.
Only recently, the question of cloning states with a fixed degree of entanglement 
has been addressed. 
The optimal operation cloning maximally entangled states of two qubits  was given in \cite{lamoureux2004},
while the more general case of non-maximally entangled states was solved in \cite{novotny2004}.

A conceptually different approach,
which may shed new light on the nature of entanglement, 
is {\em local} cloning of 
entangled states: The state is then distributed 
between two parties that may only act locally on their respective
subsystems, with or without 
classical communication.
The impossibility of locally cloning  a {\em known} entangled state has
been pointed out and used to define an entanglement measure \cite{sens}. 
Cloning of orthogonal entangled states with local operations and classical communication (LOCC), 
under the assumption of sharing an entangled blank state has been studied in \cite{martin}.

This paper addresses the problem of finding the optimal LOCC operation for cloning 
an unknown  entangled state, with a fixed degree of entanglement.
In all previous papers touching upon this problem \cite{broadcasting1997,lamoureux2004}, classical communication was 
not considered,
and 
only 
independent Bu{\v z}ek-Hillery cloners 
were applied
to each subsystem.
Here we will show that 
classical communication produces very interesting effects.

The plan of the paper is as follows. 
We restrict ourselves  to $1\to2$ cloning (optimal production of two copies from one copy) of two-qubit pure entangled states,
such that all states with the same degree of entanglement are cloned equally well,
namely with the same fidelity. The degree of entanglement is explicitly given in Sect. \ref{sec-dui}.
In the three succeeding sections, we find the optimal local  protocol for cloning two-qubit states with a given amount of entanglement, and 
show that 
only 
when the degree of entanglement is higher than a certain ``critical"  threshold, 
classical communication improves 
the cloning fidelity. In particular, this critical threshold, is obtained in Sect. \ref{subsec-tin-ka}. 
(For other examples of usefulness of 
classical communication, 
see Refs. \cite{jomey-doi, bisarga}.) 
This is, to our knowledge, the first quantum protocol for which a critical entanglement threshold for usefulness
of classical communication is demonstrated. 
In Sect. \ref{sec-chhoi},
we  show that bound entangled states with positive partial transpose (PPT) \cite{HHH_bound} are not useful as a resource for local cloning. 
We discuss our results in Sect. \ref{sec-Mithhun}.

\section{Degree of entanglement}
\label{sec-dui}

The degree of entanglement of a bipartite state
can be quantified with the help of the Schmidt decomposition. Every two-qubit state
can be written as
\[\ket{\Phi}=U_A \otimes U_B \left(\alpha \ket{00} + \sqrt{1-\alpha^2} \ket{11}\right),\] 
where $U_A$, $U_B$ are single qubit 
unitary operations, 
and $\alpha\in (0,1/\sqrt{2})$ is a 
Schmidt coefficient determining the degree of entanglement of the state \cite{Bennett-partial}.
We look for the cloning transformation that clones all states with a given $\alpha$ 
with the same fidelity.

\section{Setting up the stage: The constraints on a local cloner}

The cloning transformation  is a linear and
completely positive (CP)
 map $\mathcal{E}$
 that takes a two-qubit input state 
\[\rho_{\textrm{in}}=\ket{\Phi}\bra{\Phi} \in \mathcal{L}(\mathcal{H}_A \otimes \mathcal{H}_B)\]
to a four-qubit state 
\[\rho_{\textrm{out}}\in \mathcal{L}(\mathcal{H}_{1A} \otimes \mathcal{H}_{1B} \otimes \mathcal{H}_{2A}\otimes \mathcal{H}_{2B}).\]
(\(\mathcal{L}(\mathcal{H})\) denotes the set of all states on a Hilbert space \(\mathcal{H}\).)
The reduced density matrices of the two clones are
 $\rho_{1(2)}=\textrm{Tr}_{2A,2B(1A,1B)}(\rho_{\textrm{out}})$, where
$\textrm{Tr}_{2A,2B(1A,1B)}$ denotes tracing over $\mathcal{H}_{2A}\otimes \mathcal{H}_{2B}$ ($\mathcal{H}_{1A} \otimes \mathcal{H}_{1B}$).
In the following, we investigate symmetric cloning, i.e. the situation when $\rho_1 = \rho_2$. 
As a 
figure of merit for optimality, we adopt 
here
the so-called local fidelity
\[F= \bra{\Phi} \rho_{1(2)} \ket{\Phi},\]
 quantifying the similarity of each of the clones to the input.
This is the figure of merit which judges each clone independently, and 
is more in the spirit of original formulation of the problem of cloning than the global fidelity (see e.g. \cite{werner1998}).

A fact which is often exploited in investigations on optimal cloning is, that 
 the optimum can always be reached by a covariant operation 
\cite{werner1998,novotny2004}. This is also true in our case, and so 
we impose covariance 
with respect to local unitary operations:  
\[\mathcal{E}(U_A \otimes U_B \rho_{\textrm{in}} U_A^\dagger \otimes U_B^\dagger)
=(U_A \otimes U_B)^{\otimes 2} \mathcal{E}(\rho_{\textrm{in}}) (U_A^\dagger \otimes U_B^\dagger)^{\otimes 2}.\]
In general, with every CP map $\mathcal{E}: \mathcal{L}(\mathcal{H}) \to \mathcal{L}(\mathcal{K})$, 
one can associate a positive operator $P_\mathcal{E} \in \mathcal{L}(\mathcal{K}\otimes\mathcal{H})$, 
via the 
Jamio{\l}kowski isomorphism \cite{jamiolkowski1972}. The covariance condition of the cloning map $\mathcal{E}$ 
 can be  written using the corresponding positive operator 
$P_\mathcal{E}\in \mathcal{L}(\mathcal{H}_{1A} \otimes \mathcal{H}_{1B} 
\otimes \mathcal{H}_{2A}\otimes \mathcal{H}_{2B} \otimes \mathcal{H}_A \otimes \mathcal{H}_B)$ 
as follows \cite{ariano2001}:
\[[P_\mathcal{E},U_A \otimes U_B \otimes U_A \otimes U_B \otimes U_A^* \otimes U_B^*]=0.\]
For convenience, let us introduce an operator $\tilde{P}_\mathcal{E}$ which is the operator $P_\mathcal{E}$ but with 
different ordering of subspaces, namely:
$\tilde{P}_\mathcal{E} \in \mathcal{L}(\mathcal{H}_{1A} \otimes \mathcal{H}_{2A} \otimes \mathcal{H}_{A}\otimes 
\mathcal{H}_{1B} \otimes \mathcal{H}_{2B} \otimes \mathcal{H}_B)$. The covariance condition for $\tilde{P}_\mathcal{E}$ reads
\[[\tilde{P}_\mathcal{E},U_A \otimes U_A \otimes U_A^* \otimes U_B \otimes U_B \otimes U_B^*]=0.\]
In order to write the most general form of $\tilde{P}_\mathcal{E}$ satisfying the above condition, let us first write the most general
form of a hermitian operator $A$, satisfying 
\[[A,U_A \otimes U_A \otimes U_A^*]=0.\]
The space $\mathcal{H}_{1A} \otimes \mathcal{H}_{2A} \otimes \mathcal{H}_{A}$ can be decomposed into invariant subspaces under the action of 
$U_A \otimes U_A \otimes U_A^*$. There are two two-dimensional invariant subspaces and one four dimensional subspace \cite{was_a_footnote1}:
\(\mathcal{H}_{1A} \otimes \mathcal{H}_{2A} \otimes \mathcal{H}_{A} = \mathcal{M}_1 \oplus  \mathcal{M}_2 \oplus  \mathcal{M}_3\).
In what follows, $\mathcal{M}_1$ is the space spanned by the vectors $(\ket{011}-\ket{101})/\sqrt{2}$ and $(\ket{010}-\ket{100})/\sqrt{2}$,
$\mathcal{M}_2$ is spanned by  $(2\ket{000}+\ket{011}+\ket{101})/\sqrt{6}$ and $(\ket{010}+\ket{100}+2\ket{111})/\sqrt{6}$, while $\mathcal{M}_3$ 
by the remaining four orthogonal vectors.
Let $T_i$ be a projection operator on the subspace $\mathcal{M}_i$. Let $T_{12}$ ($T_{21}$) be an isomorphism from 
 $\mathcal{M}_1$ to $\mathcal{M}_2$ ($\mathcal{M}_2$ to $\mathcal{M}_1$). For convenience we also introduce Hermitian operators 
$T_4 = T_{12}+T_{21}$ and $T_5=iT_{12} - iT_{21}$. 
The subspaces $\mathcal{M}_1$ and  $\mathcal{M}_2$ support equivalent two-dimensional irreducible representations of $\textrm{SU(2)}$, while 
 $\mathcal{M}_3$ supports a four-dimensional irreducible representation.
Using Schur's lemma, one can show that the most general form of a Hermitian operator $A$, satisfying
covariance condition, 
has the form \cite{ariano2001}
\[A=\sum_{i=1}^5 a_i T_i,\] 
where $a_i$ are arbitrary real parameters.
Hence, the most general form of a positive operator $\tilde{P}_\mathcal{E}$, satisfying covariance condition,
reads 
\[\tilde{P}_{\mathcal{E}} = \sum_{i,j=1}^5 a_{ij} T_i \otimes T_j,\]
where we now have 25 real parameters $a_{ij}$, which have to be chosen 
such that the operator $\tilde{P}_{\mathcal{E}}$ is positive. 
Additional constraints 
come from the fact that 
the CP map corresponding to the positive operator is trace preserving. This is the case provided that
\(\textrm{Tr}_{1A,1B,2A,2B}{P_\mathcal{E}} = \openone \in \mathcal{L}(\mathcal{H}_A \otimes \mathcal{H}_B)\),
which imposes one linear equality for the $a_{ij}$:
\[a_{11} + a_{12} + 2a_{13} + a_{21}+  a_{22} + 2a_{23} + 2a_{31} + 2a_{32} +  4a_{33} = 1.\]
There is also a number of linear constraints coming from the symmetry condition on the two clones, $\rho_{1}=\rho_{2}$. 
Lastly, there is the condition that the operation should be LOCC.

The symmetry condition together with the LOCC constraint
make the problem much more difficult than the problem of optimal global cloning
with the global fidelity $F_g= \bra{\Phi}\bra{\Phi} \rho_{\textrm{out}} \ket{\Phi}\ket{\Phi}$
 as a figure of merit,
 as considered in \cite{novotny2004}, where the optimal cloning operation was found analytically \cite{guchhiey-lyang}. 
We first deal with the problem numerically, and  this leads us to an analytical solution.

\section{The optimal local cloner and the threshold for usefulness of classical communication}

\subsection{Finding the optimal local cloner}

Disregarding for the moment the LOCC constraint, 
the search for the optimal cloning operation can be written as a semidefinite program. 
In a semidefinite program, one tries to maximize a linear function of variables, while keeping certain matrices (that depend in a linear way 
on these variables) positive semidefinite.
One can also impose additional linear equalities on the variables. This is 
just 
the kind of problem we are dealing with.
The fidelity we want to maximize is linear in the parameters,
we have linear equalities (trace preservation, symmetry), and we have to keep the operator $\tilde{P}_\mathcal{E}$ positive 
semidefinite. 

Imposing the LOCC constraint is in general not easy, so we first impose a weaker constraint, namely the positivity of the partial transpose  of 
$\tilde{P}_\mathcal{E}$ with respect to the subsystem $B$ ($\mathcal{H}_{1B} \otimes \mathcal{H}_{2B} \otimes \mathcal{H}_B$) (see \cite{Lewen-Jamiol}). 
The
PPT condition is a weaker condition, because 
if an operation is LOCC it is separable, and if it is separable it satisfies PPT.
(The opposite implications do not hold \cite{HHH_bound, nlwe}.)
The PPT condition has the advantage that it can be easily incorporated into a semidefinite program. Fortunately, 
no stronger constraint than PPT will be needed, because in our case imposing the PPT constraint 
will yield an operation which will be proven later to be LOCC.

\begin{figure}[t]
\begin{center}
\includegraphics[width=0.5\textwidth]{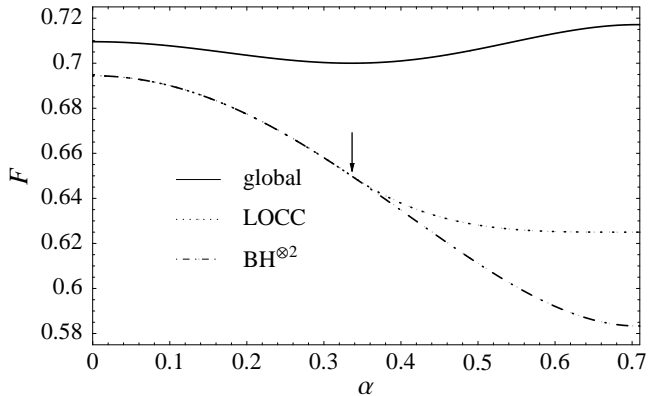}
\caption{Fidelities of three different operations performing cloning of entangled states, with a given degree of entanglement $\alpha$. 
}
 \label{fig:fidglobPPT}
\end{center}
\end{figure}

We have performed the numerics in Matlab using a package for semidefinite programing called SeDuMi. In Fig. \ref{fig:fidglobPPT}
the results of optimization are presented.

\begin{figure}[t]
\begin{center}
\includegraphics[width=0.5\textwidth]{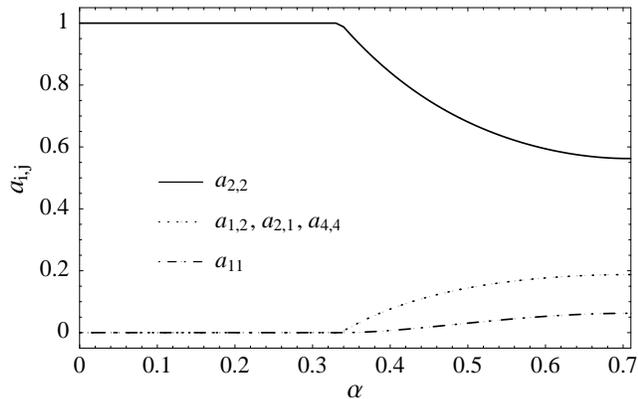}
\caption{Parameters describing the optimal LOCC operation that clones entangled states, with given degree of entanglement $\alpha$.
}
 \label{fig:paramPPT}
\end{center}
\end{figure}
The solid curve represents the optimal cloning when we do not impose a PPT constraint on
the optimization. For this operation, most
 $a_{ij}$ are zero, except for  \cite{was_a_footnote2}
\begin{eqnarray}
a_{11} &=& \frac{1}{2} - \frac{4(1-\alpha^2+\alpha^4)}{c(\alpha)},\nonumber\\
a_{22} &=& 1-a_{11}, \nonumber\\
a_{44} &=& \frac{\sqrt{a_{11}a_{22}}}{2}, \nonumber \\ 
a_{55} &=& -a_{44}, \nonumber
\end{eqnarray}
 where 
\[c(\alpha)=\sqrt{73+16\alpha^2(1-\alpha^2)(1+40\alpha^2-40\alpha^4)}.\]  
%
%
The fidelity dependence on $\alpha$ has a very similar character when optimizing local fidelity 
or global fidelity \cite{novotny2004}, with a minimum 
at 
\[\alpha=\alpha_0=\sqrt{\frac{1}{2} - \frac{\sqrt{15}}{10}} \approx .3357.\]
We found for the local fidelity
\[F=\frac{16+(1-4\alpha^2)^2-8\alpha^4+c(\alpha)}{36}.\]
For maximally entangled states ($\alpha=\sqrt{2}/2$), the 
fidelity equals $F=(5+\sqrt{13})/12$, which agrees with the result obtained in \cite{lamoureux2004}.
For product states ($\alpha=0$), the fidelity equals $F=(17+\sqrt{73})/36$. 

The dashed-dotted curve in Fig. \ref{fig:fidglobPPT} represents the fidelity of cloning when 
two optimal $1\to 2$ Bu{\v z}ek-Hillery cloners \cite{buzek1996} are applied independently to each qubit from the entangled pair
 ($\textrm{BH}^{\otimes 2}$). This is actually the optimal local cloning operation when no communication is allowed.
This can be derived writing the operator $P_{\mathcal{E}}$ as a product of two covariant local operators: $P_{A}\otimes P_{B}$,
and optimizing the fidelity, keeping in mind the trace and symmetry constraints.
The only non-zero parameter of the optimal transformation is $a_{22}=1$, and the corresponding fidelity:
\[F_{\textrm{BH}^{\otimes 2}}=\frac{25-16\alpha^2+16\alpha^4}{36}.\]

\subsection{The critical threshold of entanglement above which classical communication helps in local cloning}
\label{subsec-tin-ka}

The crux of the paper is the dotted curve in Fig. \ref{fig:fidglobPPT} which represents the fidelity of the optimal LOCC cloning.
For $\alpha$ smaller than a critical value $\alpha_0$,
the optimal LOCC operation is just $\textrm{BH}^{\otimes 2}$.
Namely, for weakly entangled states, classical communication between the parties does not help to improve the
fidelity of local cloning.
Beyond the threshold, i.e.
 for $\alpha>\alpha_0$, $\textrm{BH}^{\otimes 2}$ ceases to be the optimal
 LOCC cloning protocol.
The point of the transition 
 is indicated by an arrow. Notice that this point
 is exactly the point at which the unrestricted global cloning 
has the minimum.
For $\alpha>\alpha_0$, the corresponding non-zero parameters are given by 
\begin{eqnarray}
a_{11} &=& \frac{(1-10\alpha^2+10\alpha^4)^2}{4(1+8\alpha^2-8\alpha^4)^2}, \nonumber \\
a_{22} &=& (1-\sqrt{a_{11}})^2, \nonumber \\ 
a_{12} &=& a_{21}=a_{44}=\sqrt{a_{11}a_{22}}. \nonumber
\end{eqnarray}
The optimal LOCC fidelity for $\alpha>\alpha_0$  reads 
\[F_{\textrm{LOCC}}=\frac{3+8\alpha^2(1-\alpha^2)(2+\alpha^2-\alpha^4)}{4(1+8\alpha^2-8\alpha^4)}.\]
The optimal LOCC operation outperforms $\textrm{BH}^{\otimes 2}$ most visibly in the case of cloning of maximally entangled states, where
the optimal LOCC yields the fidelity of cloning $F=5/8=0.625$, whereas 
the $\textrm{BH}^{\otimes 2}$ operation can achieve only $F=7/12 \approx 0.583$.

In Fig. \ref{fig:paramPPT}, we present the dependence on $\alpha$, of 
the non-zero $a_{ij}$ of the optimal LOCC transformation.
For $\alpha\leq \alpha_0$, the only non-zero parameter is $a_{22}=1$, and the optimal operation is $\textrm{BH}^{\otimes 2}$. 
For $\alpha>\alpha_0$, there are five non-zero $a_{ij}$'s describing the optimal transformation, of which three are identical.
The non-differentiable behaviour of $a_{ij}$ at $\alpha=\alpha_0$
 is quite a surprising effect, as it indicates that there is a ``critical" 
threshold $\alpha_0$ for the degree of entanglement of two-qubit states,
below which  classical communication does not help in enhancing the fidelity of cloning, while above the threshold, classical 
communication does improve the cloning fidelity.

\section{Explicit protocol for implementing the optimal local cloner}

We proceed now to give the promised proof that the optimal LOCC cloning transformation that we  discussed is indeed LOCC.
Recall that we have obtained this transformation  numerically, imposing only the
PPT constraint, so in principle it could be a nonlocal operation.

Having the explicit form of the positive operator $P_\mathcal{E}$ 
that corresponds to this operation, we will first show that it is a separable operator, 
by expressing the corresponding 
 CP map $\mathcal{E}$ in terms of {\em local} Kraus operators
\(K_i\)
as 
 \[ \mathcal{E}(\rho_{\textrm{in}}) = \sum_{i=1}^8 K_i \rho_{\textrm{in}} K_i^\dagger,\quad \sum_i K_i^\dagger K_i = \openone,\]
where $K_i$ act from a two qubit input space into the tensor product of two qubit spaces of the two clones:
\begin{equation}
  K_i:\ \underbrace{\mathcal{H}_A \otimes \mathcal{H}_B}_{\textrm{input space}} 
\mapsto \underbrace{\mathcal{H}_{1A} \otimes \mathcal{H}_{1B}}_{\textrm{first clone}}\otimes
\underbrace{ \mathcal{H}_{2A} \otimes \mathcal{H}_{2B}}_{\textrm{second clone}}.
\nonumber 
\end{equation}
Additionally, the 
$K_i$ can be written as products of local Kraus operators acting on the $A$ or $B$ subsystem as 
 \[ K_i = K_{i,A} \otimes K_{i,B},\] 
where
\[ K_{i,s}:  \mathcal{H}_s \mapsto \mathcal{H}_{1s} \otimes \mathcal{H}_{2s}, \quad s = A,B.\]
%
In order to write down the $K_i$ operators in a simple manner it will prove useful to introduce four matrices $M_i$:
\begin{equation}
\begin{array}{rclrcl}
M_{1}&=&\left(\begin{array}{cc}w & 0 \\ 0 & \frac{1}{2}w-v \\ 0 &\frac{1}{2}w+v \\ 0 & 0  \end{array}\right),& 
M_{2}&=&\left(\begin{array}{cc}w & 0 \\ 0 & \frac{1}{2}w+v \\ 0 &\frac{1}{2}w-v \\ 0 & 0  \end{array}\right), \\ \\
\end{array}
\nonumber
\end{equation}
\begin{equation}
\begin{array}{rclrcl}
M_{3}&=&\left(\begin{array}{cc}0 & 0 \\ \frac{1}{2}w+v & 0 \\\frac{1}{2}w-v &0\\ 0 & w  \end{array}\right), &
M_{4}&=&\left(\begin{array}{cc}0 & 0 \\ \frac{1}{2}w-v & 0 \\\frac{1}{2}w+v &0\\ 0 & w  \end{array}\right), 
\end{array}
\nonumber
\end{equation}
where 
\[w= \frac{a_{22}^{1/4}}{\sqrt{3}}, \quad v=\frac{a_{11}^{1/4}}{2}.\]
Then
\begin{eqnarray}
K_1=\sqrt{2} M_1 \otimes M_1, &&  
K_5=\sqrt{2} M_2 \otimes M_2,  \nonumber \\
K_2=\sqrt{2} M_1 \otimes M_3,&&  
K_6=\sqrt{2} M_2 \otimes M_4,  \nonumber \\
K_3=\sqrt{2} M_3 \otimes M_1,&&  
K_7=\sqrt{2} M_4 \otimes M_2, \nonumber \\
K_4=\sqrt{2} M_3 \otimes M_3,&&  
K_8=\sqrt{2} M_4 \otimes M_4, \nonumber 
\end{eqnarray}
where the prefactor $\sqrt{2}$ is needed 
for 
trace preservation.
The LOCC protocol which realizes the above operation requires
 only one-way communication of a single classical bit and can be implemented
 as follows.
First, a local POVM (positive operator valued measurement) 
is performed on $A$:
\[\ket{\Phi}\bra{\Phi} \to \sum_{i=1}^4 M_i \otimes \openone \ket{\Phi}\bra{\Phi} M^\dagger_i \otimes \openone\]
(\(\sum_{i=1}^{4} M_i^\dagger M_i = \openone\)).
Next, a (one-bit) classical one-way communication
informs
B whether the result \(i\) of the POVM on 
$A$ was in $\{1,3\}$, or
$\{2,4\}$. In the first (second) case, \(B\) is operated on by 
using the Kraus operators
$\sqrt{2}M_{1}, \sqrt{2}M_{3}$ ($\sqrt{2}M_{2}, \sqrt{2}M_{4}$), completing the protocol.  
The equality  $M_{1}^\dagger M_{1} + M_{3}^\dagger M_{3} = \openone / 2$ (and similarly for $M_{2}, M_{4}$), 
guarantee
that the 
protocol is indeed a deterministic LOCC.

To perform the POVM at her side, Alice
uses a four dimensional local ancilla ($\mathcal{H}_{aA}$),
and
performs first a unitary on $\mathcal{H}_{1A} \otimes \mathcal{H}_{2A} \otimes \mathcal{H}_{aA}$, 
which acts as
\begin{eqnarray}
U_A \ket{00a_1} = w\ket{00}\left(\ket{a_1}+\ket{a_2}\right)+ \ket{\underbar{01}^+}\ket{a_3} +
\ket{\underbar{01}^-}\ket{a_4}, 
\nonumber \\
U_A \ket{10a_1} =\ket{\underbar{01}^-}\ket{a_1} +
\ket{\underbar{01}^+}\ket{a_2} + w\ket{11}\left(\ket{a_3}+\ket{a_4}\right), 
\nonumber 
\end{eqnarray}
where 
\[\ket{\underbar{01}^\pm} = (\frac{1}{2}w \pm v)\ket{01}+(\frac{1}{2}w \mp v)\ket{10},\]
and $\ket{a_i}$ are 
orthogonal,
and measures on \(\{|a_i\rangle\}\).
%
%
%
Bob requires only a two-dimensional ancilla ($\mathcal{H}_{aB}$),
and uses the unitaries given by
%
%
\begin{eqnarray}
U_B^{\pm} \ket{00a_1} = \sqrt{2}w\ket{00a_1}+ \sqrt{2} \ket{\underbar{01}^\pm} \ket{a_2}, 
\nonumber \\
U_B^{\pm} \ket{10a_1} =\sqrt{2} \ket{\underbar{01}^\mp}  \ket{a_1} + \sqrt{2}w\ket{11a_2}, \nonumber
\end{eqnarray}
where the upper (lower) sign
 corresponds to the use of 
$\sqrt{2}M_{1}, \sqrt{2}M_{3}$ 
($\sqrt{2}M_{2}, \sqrt{2}M_{4}$).

\section{Bound entangled states with positive partial transpose are not useful as additional resources in local cloning}
\label{sec-chhoi}

It is interesting to note 
that even though we have imposed a 
PPT 
constraint on the cloning operation, which is weaker than an LOCC constraint, we have found an LOCC 
operation to be the optimal one.
This means that there are no 
PPT 
operations that helps to improve the fidelity over 
 LOCC.
In particular, this implies that if apart from the ability to perform LOCC,
 we were given some PPT bound entangled states \cite{HHH_bound}
as additional resources, this would not help in the local cloning process
(cf., e.g. \cite{activation}).
This behaviour is different  from, e.g., the recent observations
regarding the convertibility properties of pure quantum states, where
  PPT operations can
significantly outperfom LOCC operations \cite{martin2}.

\section{Discussion}
\label{sec-Mithhun}

A local cloning machine has several potential applications. In particular, in secret sharing \cite{Marek} 
(see also \cite{byapok}), a boss requires to 
send the states 
\[\frac{1}{\sqrt{2}}(|00\rangle \pm |11\rangle), \quad \frac{1}{\sqrt{2}}(|00\rangle \pm i|11\rangle)\] to 
two subordinates.
Our local cloning machine gives bounds on the security in  a local  eavesdropping on this protocol.
Moreover, 
for a different secret sharing protocol (which we conjecture to be more secure \cite{darun-paper} than the one in \cite{Marek}), 
that uses the three quartets 
\begin{eqnarray}
\left\{\frac{1}{\sqrt{2}}(|\phi_1\rangle \pm |\phi_2\rangle),  \frac{1}{\sqrt{2}}(|\phi_3\rangle \pm |\phi_4\rangle)\right\}, \nonumber \\ 
\left\{\frac{1}{\sqrt{2}}(|\phi_1\rangle \pm |\phi_3\rangle),  \frac{1}{\sqrt{2}}(|\phi_2\rangle \pm |\phi_4\rangle)\right\}, \nonumber \\
\left\{\frac{1}{\sqrt{2}}(|\phi_1\rangle \pm |\phi_4\rangle),  \frac{1}{\sqrt{2}}(|\phi_2\rangle \pm |\phi_4\rangle)\right\}, \nonumber 
\end{eqnarray}
our local cloning machine gives the 
optimal local attack, when considering
cloning-based strategies.
Here, 
\begin{eqnarray}
|\phi_1\rangle = \frac{1}{\sqrt{2}}(|00\rangle + |11\rangle), && 
|\phi_2\rangle = \frac{-i}{\sqrt{2}}(|00\rangle - |11\rangle), \nonumber \\
|\phi_3\rangle = \frac{1}{\sqrt{2}}(|01\rangle - |10\rangle), &&
|\phi_4\rangle = \frac{-i}{\sqrt{2}}(|01\rangle + |10\rangle). \nonumber
\end{eqnarray}
%
Note that a  local attack is the realistic one for secret sharing, and, to our knowledge,
has not yet been considered \cite{darun-paper}.


In conclusion, we have found the optimal fidelity for local cloning of
pure two-qubit states, when the degree of entanglement is fixed. For
weakly entangled states, the optimal transformation is shown to  consist
of two independent optimal single-qubit cloners, a la Bu{\v z}ek-Hillery.
For states with entanglement that is higher than a certain threshold, 
communication between the parties improves the fidelity considerably.
We have thus pointed out the existence of a critical entanglement value
for the usefulness of classical communication in local cloning.
Moreover, non-local operations that are separable or 
PPT 
 do not lead to any advantage beyond local operations and classical communications. This, in particular, implies 
 that bound entangled states with positive partial transpose are not useful 
 as an additional resource for local cloning.

\begin{acknowledgments}

We acknowledge discussions with Toni Ac{\'\i}n, and support from the DFG (SFB 407, SPP 1078, 432 POL), 
the AvH
Foundation, the EC Program 
QUPRODIS, the ESF Program QUDEDIS, EU IP SCALA and SECOQC,
and the Polish Ministry of Scientific Research and Information Technology under the (solicited) grant No.
PBZ-Min-008/P03/03.

\end{acknowledgments}


\begin{thebibliography}{99}



\bibitem{pakhi} Th. Richter and W. Vogel,
Phys. Rev. Lett. \textbf{89}, 283601 (2002). 

\bibitem{lagaan} M.A. Nielsen and I.C. Chuang, \emph{Quantum Computation and Quantum Information} (Cambridge University Press, Cambridge, 2000).


\bibitem{sundar} A.S. Holevo, Probl. Pereda. Inf. \textbf{9}, 3 1973 [Probl. Inf.
Transm. \textbf{9}, 110 (1973)]; R. Josza, D. Robb, and W. Wotters, Phys. Rev. A, \textbf{49}, 668 (1994).


\bibitem{royal-bengal} P. Badzi{\c{a}}g, M. Horodecki, A. Sen(De), and
U. Sen, Phys. Rev. Lett. \textbf{91}, 117901 (2003); 
A. Sen(De), U. Sen, and M. Lewenstein, quant-ph/0505137. 



\bibitem{chandrabindu} C.H. Bennett and S.J. Wiesner, 
Phys. Rev. Lett. \textbf{69}, 2881 (1992). 

\bibitem{bisarga} D. Bru\ss, G.M. D'Ariano, M. Lewenstein, C. Macchiavello, A. Sen(De), and U. Sen,
Phys. Rev. Lett. \textbf{93}, 210501 (2004).


\bibitem{wootters} W.K. Wootters and W.H. Zurek, Nature (London) \textbf{299}, 802 (1982); D. Dieks, Phys. Lett. \textbf{92A}, 271 (1982).

 \bibitem{buzek1996}  
V. Bu{\v z}ek and M. Hillery,
Phys. Rev. A \textbf{54}, 1844 (1996). 


\bibitem{cloning} 
 D. Bru{\ss}, 
D.P. DiVincenzo, A. Ekert, C.A. Fuchs, C. Macchiavello, and J.A. Smolin,
Phys. Rev. A \textbf{57}, 2368 (1998); R. Werner, \emph{ibid.} {\bf 58}, 1827 (1998).


\bibitem{werner1998} M. Keyl and R.F. Werner,
J. Math. Phys. \textbf{40}, 3283 (1999).


\bibitem{cloninf} N.J. Cerf, A. Ipe, and X. Rottenberg,
Phys. Rev. Lett. \textbf{85}, 1754 (2000).

\bibitem{lamoureux2004}
        L.-P. Lamoureux, P. Navez, J.
                  Fiur\'{a}\v{s}ek, and N. J. Cerf,
        Phys. Rev. A
        \textbf{69}, 040301 (R), 
        (2004);
        E. Karpov, P. Navez, and N. J. Cerf, Phys. Rev. A \textbf{72}, 042314 (2005).


\bibitem{novotny2004}
J. Novotn\'{y}, G. Alber, and I. Jex, Phys. Rev. A \textbf{71}, 042332 (2005).

\bibitem{sens}  M. Horodecki, A. Sen(De), and U. Sen,
Phys. Rev. A \textbf{70}, 052326 (2004).

\bibitem{martin}  S. Ghosh, G. Kar, and A. Roy,
Phys. Rev. A \textbf{69}, 052312 (2004); F. Anselmi, A. Chefles, and M. B. Plenio,
New. J. Phys. \textbf{6}, 164 (2004).

\bibitem{broadcasting1997}
V. Bu{\v z}ek, V. Vedral, M.B. Plenio, P.L. Knight, M. Hillery, Phys. Rev A \textbf{55}, 3327 (1997).


\bibitem{jomey-doi} C.H. Bennett,
G. Brassard, C. Crepeau, R. Jozsa, A. Peres, and W.K. Wootters,
Phys. Rev. Lett. \textbf{70}, 1895 (1993); C.H. Bennett, 
G. Brassard, S. Popescu, B. Schumacher, J.A. Smolin, W.K. Wootters,
\emph{ibid.} \textbf{76}, 722 (1996); C.H. Bennett, D.P. DiVincenzo, J.A. Smolin, and W.K. Wootters,
Phys. Rev. A \textbf{54}, 3824 (1996); J. Walgate, A.J. Short, L. Hardy, and V. Vedral,
Phys. Rev. Lett. \textbf{85}, 4972 (2000). 

\bibitem{HHH_bound} P. Horodecki, Phys. Lett. A \textbf{232}, 333 (1997); M. Horodecki, P. Horodecki, and R. Horodecki,
Phys. Rev. Lett. \textbf{80}, 5239 (1998).



\bibitem{Bennett-partial} C.H. Bennett, H.J. Bernstein, S. Popescu, and B. Schumacher,
Phys. Rev. A \textbf{53}, 2046 (1996).



\bibitem{jamiolkowski1972}
 A. Jamio{\l}kowski,
Rep. Math. Phys.
 \textbf{3},
 275
 (1972).

\bibitem{ariano2001}  G.M. D'Ariano and P. Lo Presti, 
Phys. Rev. A \textbf{64}, 042308 (2001).

\bibitem{was_a_footnote1} 
This is like adding three spins 
$1/2$ and looking for subspaces with fixed total angular momentum. The difference is that the third representation of 
$\textrm{SU}(2)$ is complex conjugated,
which 
requires 
the renaming 
$\ket{0}\to \ket{1}$, $\ket{1}\to -\ket{0}$ in the third subspace $\mathcal{H}_A$.


\bibitem{guchhiey-lyang} 
For the global fidelity \(F_g\), classical communication turns out to be useless, and 
two independent Bu{\v z}ek-Hillery cloners  \cite{buzek1996} are optimal. 


\bibitem{Lewen-Jamiol} J.I. Cirac, W. D{\" u}r, B. Kraus, and M. Lewenstein,
Phys. Rev. Lett. \textbf{86}, 544 (2001).





\bibitem{nlwe} C.H. Bennett, 
D.P. DiVincenzo, C.A. Fuchs, T. Mor, E. Rains, P.W. Shor, J.A. Smolin, and W.K. Wootters,
Phys. Rev. A \textbf{59}, 1070 (1999).


\bibitem{was_a_footnote2} 
Although we did numerics, we were able to obtain analytical results, 
as the numerical calculations indicated which \(a_{ij}\) can be put to zero, and with the rest, we could deal symbolically.

\bibitem{activation} P. Horodecki, M. Horodecki, and R. Horodecki,
Phys. Rev. Lett.  \textbf{82}, 1056 (1999);  P.W. Shor, J.A. Smolin, and B.M. Terhal,
\emph{ibid.} \textbf{86}, 2681 (2001);
T. Eggeling, K.G.H. Vollbrecht, R.F. Werner, and M.M. Wolf,
\emph{ibid.} \textbf{87}, 257902 (2001).


\bibitem{martin2} S. Ishizaka, Phys. Rev. Lett. \textbf{93}, 190501 (2004);
 S. Ishizaka and M. B. Plenio, quant-ph/0412193.


\bibitem{Marek} M. \.Zukowski, A. Zeilinger, M. Horne, and H. Weinfurter, Acta Phys. Pol. A \textbf{93}, 187 (1998);
M. Hillery, V. Bu{\v z}ek, and A. Berthiaume, Phys. Rev. A \textbf{59}, 1829 (1999).


\bibitem{byapok}  V. Scarani and  N. Gisin, Phys. Rev. Lett. \textbf{87},  117901 (2001);
 V. Scarani and  N. Gisin, Phys. Rev. A \textbf{65}, 012311 (2002);
  A. Sen(De), U. Sen, and M. \.Zukowski,
Phys. Rev. A \textbf{68}, 032309 (2003).


\bibitem{darun-paper} M. Lewenstein \emph{et al.}, in preparation.



\end{thebibliography}


\end{document}